\documentclass[12pt]{elsart}
\usepackage{graphicx,psfrag}
\def\be{\begin{eqnarray}}
\def\ee{\end{eqnarray}}
\def\bc{\begin{center}}
\def\ec{\end{center}}
\def\om{\omega}

\def\prt{\partial}

\def\rmd{{\rm d}}
\begin{document}
\begin{frontmatter}

\title{Level crossing of particle-hole and mesonic modes in eta mesic nuclei}

\author[YITP]{D. Jido},
\author[UMN,YITP,GSI]{E.E. Kolomeitsev},
\author[RCNP]{H. Nagahiro},
\author[Nara]{S. Hirenzaki}

\address[YITP]{Yukawa Institute for Theoretical Physics, Kyoto University, Kyoto 606-8502, Japan}
\address[UMN]{School of Physics and Astronomy, University of Minnesota, 116 Church Str SE, Minneapolis 55455, USA}
\address[GSI]{Gesellschaft f\"ur Schwerionenforschung (GSI), Planck Str. 1, 64291  Darmstadt, Germany}
\address[RCNP]{Research Center for Nuclear Physics (RCNP), Osaka University, Ibaraki, Osaka 567-0047, Japan}
\address[Nara]{Department of Physics, Nara Women's University, Nara 630-8506, Japan}

\begin{abstract}
We study eta meson properties in the infinite nuclear matter and
in atomic nuclei with an emphasis on  effects of the eta
coupling to $N^*(1535)$--nucleon-hole modes. The $N^*(1535)$
resonance, which dominates the low-energy eta-nucleon scattering,
can be seen as a chiral partner of the nucleon. The change of the
chiral mass gap between the $N^*$ and the nucleon in a nuclear
medium has an impact on the properties of the eta-nucleus system.
If the $N^*$-nucleon mass gap decreases with a density increase
(chiral symmetry restoration) the calculations show the existence
of the resonance state at the energy about $60$~MeV and two bound eta-nucleus states with the binding
energies about $-80$~MeV.
These states can have strong effect on predicted cross sections of
the  $^{12}{\rm C}(\gamma,p)^{11}{\rm B}$ reaction with eta-meson production.

\end{abstract}


\end{frontmatter}
\section{Introduction}
The eta bound states in nuclei were theoretically predicted by
Haider and Liu~\cite{Haider:1986sa}. This prediction relied on the
first estimation of the $\eta N$ interaction just made that time
by Bhalerao and Liu~\cite{NNeta2}, who obtained an attractive
$\eta N$ scattering length. Being electrically neutral the eta
meson can be bound only by strong interactions. Simultaneously, the
strong interactions induce nuclear absorption of the bound eta
meson, which wave function has to have large overlap with a
nucleus. As the result the position and the width of the bound
level are determined by a complicated interplay of the nuclear density distribution
and the strength of the in-medium $\eta N$
interaction. Therefore the experimental proof of the existence of
eta bound states could provide information on modification of
meson properties in nuclear matter complimentary to that obtained from
pionic and kaonic atoms~\cite{FriedmanGal}. Particularly, the
existence of such states in light nuclei would set important
constraints~\cite{Sibirtsev04}. Currently, various theoretical
calculations predict the natural widths of  eta bound states to be
larger than the level
spacing~\cite{Kohno:1989wn,col91,Hayano:1998sy,Tsushima:1998qw,GarciaRecio:2002cu,Jido:2002yb,Nagahiro:2003iv,Nagahiro:2005gf}.
This could explain problems in experimental identification of the eta bound
states.

The first experimental search for eta bound states in ($\pi^+,p$)
reactions~\cite{Chrien:1988gn} gave a negative result as being
aimed only at the narrow states as it was proposed in
Ref.~\cite{Haider:1986sa}. At the same time the measured excitation
functions for reactions $d\,(p,^{3}{\rm He})\,
\eta$~\cite{Berger88} and $^{18}{\rm O}\,(\pi^+,\pi^-)\,^{18}{\rm
Ne}$~\cite{Johnson:1993zy} showed some enhancement near the $\eta$
production threshold which could be a hint for a bound
state. Sokol and Triasuchev argued in \cite{Sokol91} that the eta
bound state can be more easily identified if one tags on the
$N(1535)$ resonance in the final state by observing its $\pi^{+}
n$ decay. Such an identification method was realized in
experiments at the Lebedev Physical Institute studying
$(\gamma,p)$ reaction on the $^{12}{\rm C}$ target. The observed
enhancement of $\pi^{+} n$ production occurring when the photon
energy exceeds the eta production threshold can be interpreted as
a signal of $\eta$-nucleus formation~\cite{Sokol:1998ua}. Observation of an eta bound
state in $^{3}$He with the binding energy $(-4.4 \pm 4.2)$~MeV and
the full width $(25.6\pm 6.1)$~MeV in photoproduction reactions at
MAMI was reported in~\cite{Pfeiffer:2003zd}. The bound state was
identified by registration of two decay modes with an explicit eta in the
final state and with $\pi^{0}p$. Interpretation of these
experimental results in terms of the eta bound state was, however,
questioned in~\cite{Hanhart05}. A new support for $\eta ^{3}{\rm
He}$ bound state arises from the recent study of $d\, p\to
^{3}{\rm He}\, \eta$ reaction at COSY~\cite{Wilkin07}. Other
experiments have been proposed for searching eta bound
states~\cite{Baskov:2003vr,Anikina:2004ps,Jha:2007mr}.

Proton pickup processes are powerful experimental tools for
studying light mesons in nuclei. They allow for the recoilless
creation of a meson inside a nucleus by fine tuning the energy of
an incident particle. It is well-known, e.g., that the
(d,$^{3}$He) reactions on nuclear targets lead to formation and
identification of the deeply bound pionic
atoms~\cite{NPA530etc,PRC62,PRL88,PLB514}. The $(\gamma,p)$ and
$(\pi,p)$ reactions have also been proposed as good experimental
tools for formation of eta-nucleus systems. In the recoilless
production, the eta meson is created almost at rest and may be
absorbed within the nucleus. Although the eta meson is not
directly observed in the final states, we can extract properties
of the eta-nucleus system by measuring of the production cross
section as a functions of the momentum of the emitted
particle~\cite{Hayano:1998sy,Jido:2002yb,Nagahiro:2003iv,Nagahiro:2005gf}.
Thereby one has access to the energies corresponding to eta mesons
bound in nuclei.
This is one of the differences from the eta production reactions on nuclear targets, where the eta meson emitted from the nucleus is observed~\cite{Yorita:2000bu,Maruyama:2002fz,Lehr:2003km}.

The prominent feature of the eta meson is that the eta-nucleon
system couples dominantly to the $N^{*}(1535)$ $(S_{11})$
resonance at threshold energies~\cite{eta1}. The $\eta NN$ coupling
is, on the contrary, much smaller than $\eta N N^*$ and $\pi NN$
ones~\cite{Tiator94} and is basically irrelevant for
analyses of the $NN$ scattering and the threshold $\eta$ production
on a nucleon~\cite{NNeta2,NNeta1}.
Therefore  the low-momentum eta meson would dominantly excite $N^{*}$--nucleon-hole ($N^{*}-h$) states in
nuclear medium. The difference between the eta mass and the $N^*-N$ mass
gap is just 50 MeV, which is smaller than the $N^*$ width
$\Gamma_{N^*}\simeq 75$~MeV and is comparable with the depth of
the attractive potential acting on the nucleon in a nucleus, $\sim 50-60$~MeV.
Hence, one could expect the eta meson nuclear dynamics to be
sensitive to modifications of nucleon and $N^*$ properties in
medium.

The question about in-medium dynamics of the $N(1535)$ resonance
addresses an interesting aspects of the chiral symmetry of strong
interactions. Being the lowest lying baryon with the opposite
parity to the nucleon, the $N(1535)$ can be viewed as its chiral
partner~\cite{Christos:1982kc,DK,Hatsuda:1988mv,Jido:1996ia,cdmod}.
Analogously one considers pairs $(\sigma, \pi)$ and $(\rho,
a_{1})$. If now the chiral symmetry is getting restored with an
increasing baryon density the parity partners will become
degenerate. This picture has several interesting
phenomenological consequences~\cite{Christos:1982kc,DK,cdmod}.
The important implication for the eta meson physics is that
the $N-N^*$ mass gap decreases in nuclear medium and at some
density goes below the $\eta$ mass~\cite{Kim:1998up,Hatsuda:1988mv}.
This changes the sign of the $\eta$ meson optical potential making
it repulsive at saturation density as found
in~\cite{Jido:2002yb,Nagahiro:2003iv,Nagahiro:2005gf}.

The parity-doublet concept corresponds to a linear realization of
chiral symmetry in terms of hadrons. In a non-linear realization
the parity partner is not present in the Lagrangian as an
independent degree of freedom but rather manifests itself as a
pole of the scattering amplitude generated dynamically by coupled
channels. So, the $N^*(1535)$ can be viewed as a $K\Sigma$ and
$K\Lambda$ quasi-bound
state~\cite{Kaiser:1995cy,kww97,Lutz:2001yb,iov02}. In general the
dynamically generated states could be very sensitive to in medium
modifications of the meson-baryon loops they are made of. For
instance the dynamically generated $\Lambda(1405)$ resonance would
dissolve or become very broad in nuclear
medium~\cite{Waas:1997pe,Ramos:1999ku,Lutz:2001dq,Lutz:2007bh}. In the case of
the $N^*(1535)$ the in-medium modification of kaon-hyperon loops
is expected to be smaller~\cite{Waas:1997pe,Inoue:2002xw} since
there is no Pauli blocking for hyperons and $K$ mesons  interact
rather weakly with nucleons. However, even in this case both
hyperons and nucleons can couple to an attractive scalar nuclear
mean field, which is responsible for a decrease of baryon masses
in nuclear matter. Thus, the shift of the $N^*$ with respect to the
nucleon depends on how strongly hyperons and nucleons couple to the
scalar field. If the couplings are equal then the $N$ and $N^*$
masses decrease roughly in the same manner that keeps the $N^*-N$
mass gap constant. It is known, however, from physics of
hypernuclei that the hyperons are bound weaker than  nucleons.
This could lead to an increase the $N^*-N$ mass gap in the nuclear
medium.

As discussed above, these two
different pictures for $N(1535)$ imply distinct
phenomenological consequences for eta meson dynamics in the nuclear matter~\cite{Nagahiro:2003iv,Nagahiro:2005gf}.
In the parity-doublet model the eta optical potential is
shallower, that reduces the binding~\cite{Jido:2002yb,Nagahiro:2003iv,Nagahiro:2005gf}.

Typically, calculations done in previous works operated with
energy-indepen\-dent optical potentials. In the case when $N^*-N$ gap
becomes close to or smaller than the eta mass it is not a reliable
approximation anymore, and the energy dependence must be taken
into account explicitly. The eta meson quantum numbers can be
carried now not only by the eta-particle modes but also by the
$N^*$--nucleon-hole modes. In the present paper we study
the propagation of these modes in nuclear matter and finite
nuclei. We discuss a possibility of the formation of an ``eta
nucleus" by $N^*-h$ modes\footnote{Similar interpretation of
the ``eta nucleus" in terms of both eta-particle and resonance-hole modes
was proposed in~\cite{Sokol91}.}.

In Section 2 we construct the eta meson Green function in nuclear matter
exploiting the $N^{*}$ dominance in the $\eta N$ scattering.
We calculate the spectrum of eta excitations in the infinite
nuclear system and analyze how the eta
spectral density changes with the nuclear matter density. Section 3 is devoted to
eta excitations in finite nuclei. First we discuss the spectral function
convoluted with the density distribution in $^{11}$B. This gives
an
approximate view on how the nucleus would respond to the external
source with the eta quantum numbers. For better understanding the
observed bump structures we solved the Lippmann-Schwinger equation
for the eta-nucleus scattering amplitude in the momentum representation
and identify poles of the amplitude on the complex energy plane.
In Section 4 we calculate missing mass spectra in the
$(\gamma,p)$ reaction on $^{12}$C target associated with the eta meson formation.

\section{Eta meson in nuclear matter}

\subsection{Eta self-energy in nuclear matter}

In this section we investigate the eta meson in infinite uniform nuclear medium.
The propagation of the eta meson is described in terms of the in-medium Green function
with the self-energy $\Pi_{\eta}(\omega,k; \rho)$
\be
G_\eta(\om,k;\rho)=\frac{1}{\om^2-k^2-m_\eta^2-\Pi_\eta(\om,k;\rho)+i \epsilon}\,.
\label{propeta}
\ee
Here $\omega$ and $k$ denote the eta energy and momentum, $\rho$ is the
nuclear density, and $m_{\eta}$ is the vacuum eta mass. We
consider the isospin symmetrical nuclear matter.

\begin{figure}
\psfrag{eta}{$\eta$}
\psfrag{Ns}{$N^{*}$}
\centerline{\includegraphics[width=5cm]{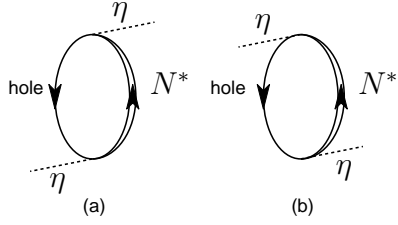}}
\caption{Diagrams for the $\eta$ self-energy in the nuclear matter. (a) and (b) are the
direct and crossing diagrams, respectively. } \label{fig:etaNNs}
\end{figure}

The eta nucleon interaction is dominated by the $N^*$ resonance
which contributes to the s-wave scattering amplitude. The higher
partial waves are less important for slow $\eta$ mesons created in
reactions at recoilless conditions. Thus,
the main contribution to the eta self-energy is due to the
excitations of $N^*$-nucleon-hole modes as shown in
Fig.~\ref{fig:etaNNs}. Exactly these diagrams induce
the strong energy dependence of the self-energy.

The $\eta N N^*$ coupling is defined by the Lagrangian
\be
\mathcal{L}_{\eta N N^*}=g\,\Big(\overline{N}\,N^*+\overline{N^*}\,
N\Big)\,\eta\,,
\label{lag}
\ee
where the coupling $g$ can be estimated as $g\simeq 2$ from the vacuum $N^*\to \eta N$ decay
width
\be
\Gamma_{N^*\to \eta N}=g^2\frac{(m_{N^{*}}+m_N)^2-m_\eta^2}{16\,\pi\,m_{N^{*}}^3}\,
\sqrt{(m_{N^{*}}^2-m_N^2-m_\eta^2)^2-4\,m_N^2\,m_\eta^2}
\,,\quad
\nonumber
\ee
with the vacuum values, $\Gamma_{N^*\to \eta N}\simeq 75~{\rm MeV}$, $m_{N}=939~{\rm MeV}$, $m_{N^{*}}=1535~{\rm MeV}$ and $m_{\eta}=547~{\rm MeV}$.

In the gas approximation the self-energy depicted in Fig.~\ref{fig:etaNNs}
reads
\be
\Pi_\eta(\om,k;\rho) &=& \frac{g^2
\rho}{\om-(m_{N^*}^*(\rho)-m_N^*(\rho))-\frac{k^2}{2\,m^*_{N^{*}}}+i\,\Gamma_{N^{*}}(\omega;\rho)/2}
\nonumber\\ &&  + \frac{g^2 \rho
}{-\om-(m^{*}_{N^*}(\rho)-m^{*}_N(\rho))-\frac{k^2}{2\,m^*_{N^{*}}}}\,,
\label{polop} \ee
where the first and second terms in
Eq.~(\ref{polop}) correspond to the diagrams (a) and (b) in
Fig.~\ref{fig:etaNNs}, respectively.
We neglect here momenta of the nucleon and the $N^*$.
The linear density dependence in the numerator of Eq.~(\ref{polop}) is
obtained by the Fermi momentum integration over the nucleon hole state.
The density-dependent in-medium masses of the $N^{*}$ and $N$ are
denoted as $m^{*}_{N^{*}}(\rho)$ and $m^{*}_{N}(\rho)$, respectively.
$\Gamma_{N^{*}}(\omega;\rho)$ is the in-medium $N^{*}$
width determined by the $N^*$ self-energy, $\Gamma_{N^{*}}(\omega;\rho)=
-2\, \Im\Sigma_{N^*}(m_N^*(\rho)+\om;\rho)$\,.
There is no width in the second term in Eq.~(\ref{polop}) since
$\Im\Sigma_{N^*}(m_N^*(\rho)-\om;\rho)=0$.

Below we use the chiral doublet model~\cite{cdmod} for the medium
modification of the $N$ and $N^{*}$ masses. In this model,
the mass difference of the nucleon and $N^{*}$ is
determined by the chiral condensate $\langle \sigma \rangle$ as
\begin{equation}
m^*_{N^*}(\rho)-m^*_N(\rho)=A \langle \sigma \rangle
\label{eq:massdif}
\end{equation}
with the parameter $A=(m_N-m_{N^*}) / \langle \sigma \rangle_{0}$, where $\langle \sigma \rangle_{0}$
is the vacuum chiral condensate.
Assuming a partial restoration of chiral symmetry in the nuclear medium and
parameterizing of the sigma condensate as a function of the nuclear density~\cite{ref:PRL82},
$\langle \sigma \rangle = (1-C\rho/\rho_0) \langle \sigma \rangle_0$, we obtain
the in-medium mass difference
\begin{equation}
m^*_{N^*}(\rho) - m^*_N(\rho)= (m_{N^{*}}-m_{N}) (1-C\,\rho/\rho_0) \ .
\label{cdmod}
\end{equation}
The parameter $C$ represents the strength of chiral restoration at the normal density and its
reasonable values are in the range of $C=0.1$--$0.3$. The value of the $C$ parameter can be
evaluated within the chiral doublet model providing $C \simeq 0.22$, which is read off the density
dependence of the chiral condensate calculated in Ref.~\cite{Kim:1998up}. In this paper we use
$C=0.2$ \cite{Jido:2002yb,Nagahiro:2003iv,Nagahiro:2005gf}.
The mass
of the eta meson is assumed to be constant in the nuclear medium
\cite{Costa:2002gk,Nagahiro:2006dr} (no scalar mean-field potential).

\begin{figure}
\centerline{\includegraphics[width=0.9\textwidth]{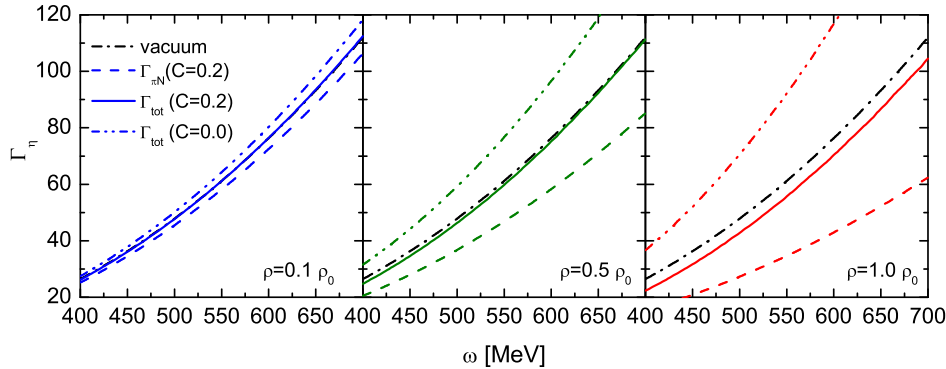}}
\caption{Energy dependence of the $N^*$ widths: the partial width
$\Gamma_{N^*\to \pi N}$ (dashed lines) and the total width
$\Gamma_{N^*}\approx \Gamma_{N^*\to \pi N}+ \Gamma_{N^*N\to
NN\pi}$ calculated for $C=0.2$ (solid lines) and $C=0.0$
(dash-dot-dot lines) are shown for different values of the nucleon
densities. Dash-dot lines show the vacuum $\Gamma_{N^*\to \pi N}$
width. For $C=0.0$ the partial width $\Gamma_{N^*\to \pi N}$
coincides with the vacuum one. } \label{fig:nswidth}
\end{figure}

There are several contributions to the imaginary part of the $N^*$ self-energy,
among which the main modes
are the pionic decay, $N^{*} \rightarrow \pi N$, and the two-body
absorption,  $N^{*}N \rightarrow \pi NN$ \cite{col91,Jido:2002yb}.
The decay process $N^*\to \eta N$ is forbidden in nuclear matter
due to the Pauli blocking. The width of $N^{*} \rightarrow \pi N$
reaction calculated in Ref.~\cite{Kim:1998up} within an improved
parity-doubled model is depicted in Fig.~\ref{fig:nswidth} (dashed
lines) as a function of the energy for different nucleon densities.
The steep rise of the width is due to the
pseudoscalar $\pi N^*N$ coupling motivated by the chiral symmetry.
For $C=0.2$ this partial width is decreasing with a growing density
as the chiral symmetry is getting restored and pions decouple from
baryons. For $C=0.0$ the width $\Gamma_{N^*\to \pi  N}$ does not depend on
the density.
The $N^{*}N \rightarrow \pi NN$ reaction has been calculated in
Ref.~\cite{Jido:2002yb}. The total $N^*$ width is depicted in Fig.~\ref{fig:nswidth}
by solid lines for $C=0.2$ and by dash-dot-dot lines for $C=0.0$.
The analysis done in this section for the case of the
homogeneous nuclear matter shows that, in principle, it is sufficient to take
density and energy independent value for the total width $\Gamma_{N^*}$
given by the vacuum value $\Gamma_{N^{*} \rightarrow \pi N} =
75$~MeV.  Later in the calculation of the
formation spectra of the eta-nucleus system, we will use
more realistic $N^{*}$ decay widths reported in
Refs.~\cite{Jido:2002yb,Nagahiro:2003iv,Nagahiro}.

We should verify how our model describes the low energy $\eta N$
interaction. The $\eta N$ scattering amplitude can be related to
the $\eta$ self-energy as
$T_{\eta N}(\om,k)=-\partial\Pi_{\eta}(\omega,k;\rho)/\partial
\rho\Big|_{\rho=0}$\,.
Using this relation, we calculate the $\eta N$ scattering length
\be
a_{\eta N} =\frac{T_{\eta
N}(m_\eta,0)}{4\,\pi\,(1+m_\eta/m_N)}\simeq 0.55 + i\,0.39~{\rm
fm}\,.
\ee
Note that $\Gamma_{\eta N}(\om=m_\eta)=0$, since the
$N^{*} \rightarrow \eta N$ channel is closed at the threshold. The
values reported in the literature are summarized in
Table~\ref{tab:scatleng}. Our value of the scattering length is
consistent with the other model calculations.

\begin{table}
\begin{tabular}{c|cccc}
$a_{\eta N}$ [fm] & $0.55 + i\,0.39$ & $0.20+i\,0.26$ & $0.26+i\,0.25$ & $0.27+i\,0.22$ \\
Ref. &this work & \cite{kww97} &\cite{iov02} &\cite{NNeta2} \\
\hline
$a_{\eta N}$ [fm] & $0.51+i\,0.21$ & $0.55\pm 0.02+i\,0.3$ & $0.88+i\,0.27$ & $0.98+i\,0.37$ \\
Ref. &\cite{sfn95} &\cite{w93} & \cite{bss95}& \cite{asy92}
\end{tabular}
\caption{\label{tab:scatleng}
Summary of the values of the $\eta N$ scattering length. }
\end{table}

With the positive $\eta N$ scattering length, the $\eta N$
interaction is attractive at the threshold. Thus, the
self-energy of the eta meson is also attractive at low densities and $\omega\simeq
m_{\eta}$. This attraction is due to the
$N^{*}$ resonance above the $\eta N$ threshold. Indeed, since the
vacuum masses of $\eta$, $N$ and $N^{*}$ satisfy $m_{\eta} <
m_{N^{*}} -m_{N}$, the real part of the self-energy has a negative
value at $\omega \simeq m_{\eta}$. At moderate densities when
medium effects for the $N$ and $N^{*}$ are not so strong and the
inequality  $m_{\eta} < m_{N^{*}}(\rho) -m_{N}(\rho)$ still holds,
the optical potential of the eta meson remains attractive
for $\omega \simeq m_{\eta}$. If some medium
effects and/or a density increase bring the $N^{*}-h$ state down
below the eta state, so that $m_{\eta} > m_{N^{*}}(\rho)
-m_{N}(\rho) $\,, the real part of the self-energy turns to be
positive and the optical potential of the eta meson becomes
repulsive~\cite{Jido:2002yb,Nagahiro:2003iv,Nagahiro:2005gf}.
Therefore, the order of mesonic and particle-hole levels determines
the sign of the self-energy, once two modes couple. Furthermore,
the level difference between these modes is only about 50 MeV.
Hence, the self-energy of the eta meson is very sensitive to the
in-medium properties of the $N$ and $N^{*}$ masses. In addition,
the level coupling brings strong energy dependence to the
self-energy. Therefore it may be wrong to discuss
the eta meson in nuclear medium in terms of an energy-independent
optical potential. In the next subsection, we discuss the spectrum
of eta modes propagating in nuclear medium taking into account
this energy dependence.

The above discussion is quite general and does not depend on
details of the mechanism of the $N$ and $N^*$ mass reduction.

\subsection{Eta spectrum in the nuclear matter}

In this section we investigate the spectral function of the eta meson
in the infinite nuclear medium.

\begin{figure}
\centerline{
\parbox{4.7cm}{\includegraphics[width=4.7cm]{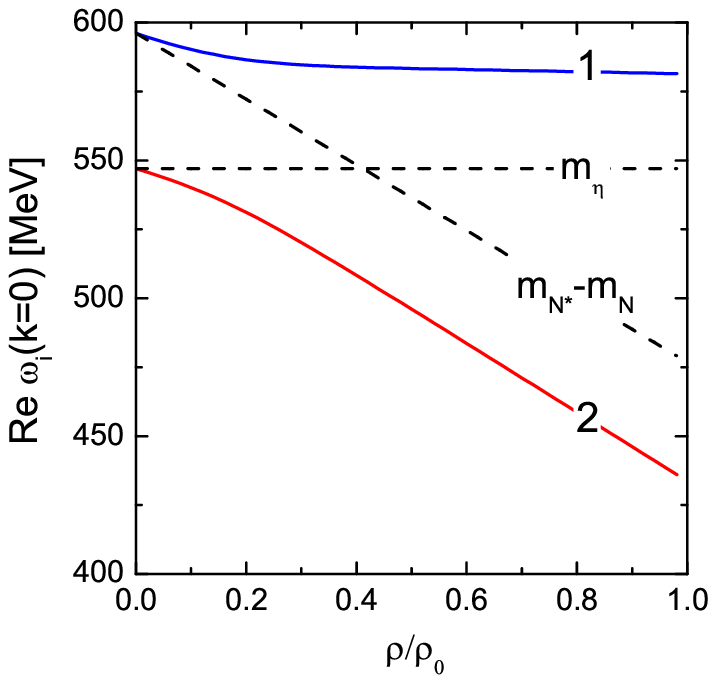}}
\parbox{4.7cm}{\includegraphics[width=4.7cm]{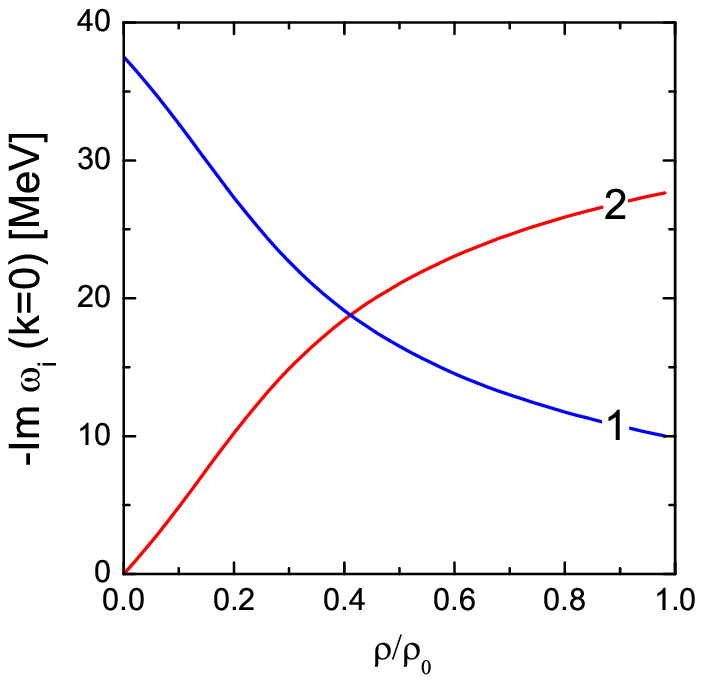}}
\parbox{4.7cm}{\includegraphics[width=4.7cm]{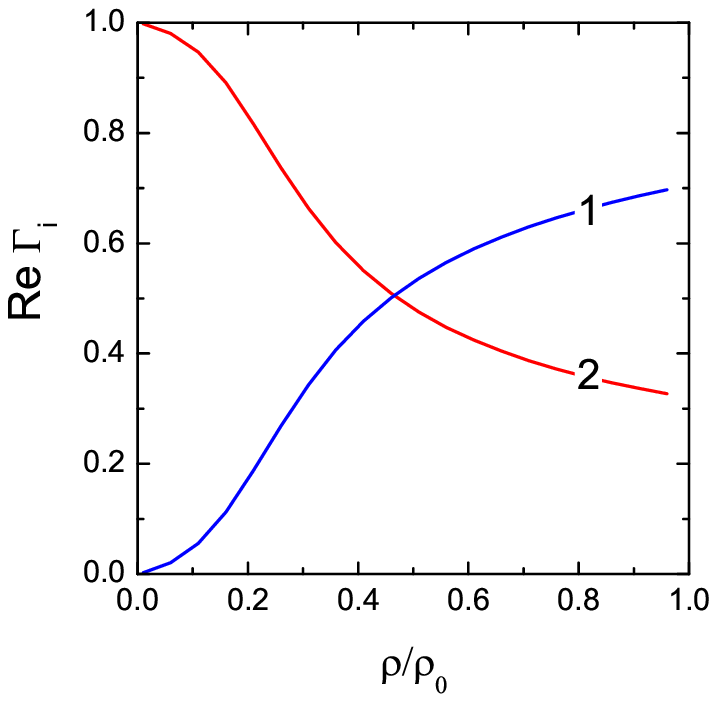}}
} \caption{ The real (left panel) and imaginary (middle panel)
parts of the solutions, $\om_i$, of eq.~(\ref{dyson1}) and the
occupations factors (\ref{occupf}) (right panel) for vanishing
momentum as functions of the baryon density. The dashed lines on
the left panel denote the eta and $N^{*}$-$h$ modes without the
channel mixing ($g=0$). Calculations are done for $C=0.2$. }
\label{fig:solk0}
\end{figure}

The full in-medium propagator of the eta meson (\ref{propeta}) can be
decomposed as
\begin{equation}
   G_{\eta}(\omega,k;\rho) = \sum_{i} \frac{\Gamma_{i}(k;\rho)}{\omega-\omega_{i}(k;\rho)} \ ,
\end{equation}
where $\omega_{i}$'s are complex numbers being the solutions of
the Dyson equation $G_\eta^{-1}(\om,k;\rho)=0$ with the
self-energy~(\ref{polop}):
\be
\om^2-k^2-m_\eta^2-g^2\,\rho\,\frac{2\,\om_R(k;\rho)
-i\,\Gamma_{N^{*}}/2}{[\om-\om_R(k;\rho)+i\,\Gamma_{N^{*}}/2]\,[\om+\om_R(k;\rho)]}
=0\,.
\label{dyson1}
\ee
We denote here $\omega_{R}(k;\rho)= m^*_{N^{*}}(\rho) - m^*_{N}(\rho) + k^{2}/(2 m^*_{N^{*}}(\rho))$.
This equation has two complex solutions with a positive real
part, $\om_i(k)$\,, $i=1,2$.
These solutions describe two branches\footnote{For the first time the two-branch spectrum of eta mesons in nuclear matter
has been discussed in Ref.~\cite{Sauermann1993}.} of the
$\eta$-meson spectrum in nuclear matter\footnote{In general case the assignment of different solutions
of the Dyson equation to the in-medium branches of particles and anti-particles should be done according to the
sign of the real parts of the residua, cf.~\cite{Migdal78} }.
They are shown in Fig.~\ref{fig:solk0} by solid lines for $k=0$ as
functions of the baryon density.
With the vanishing coupling constant $g$ the eta-meson mode and
$N^*$-nucleon-hole mode are decoupled and only the mesonic mode
can carry the $\eta$ quantum numbers. This case is depicted by the
dashed lines on the left panel in Fig.~\ref{fig:solk0}. The modes cross each other at
$\rho\simeq 0.4\,\rho_0$ for $k=0$\,. The finite value of the coupling $g$
leads to the avoided level crossing and the redistribution of the eta
quantum numbers between both branches.

\begin{figure}
\centerline{
\parbox{4.7cm}{\includegraphics[width=4.7cm]{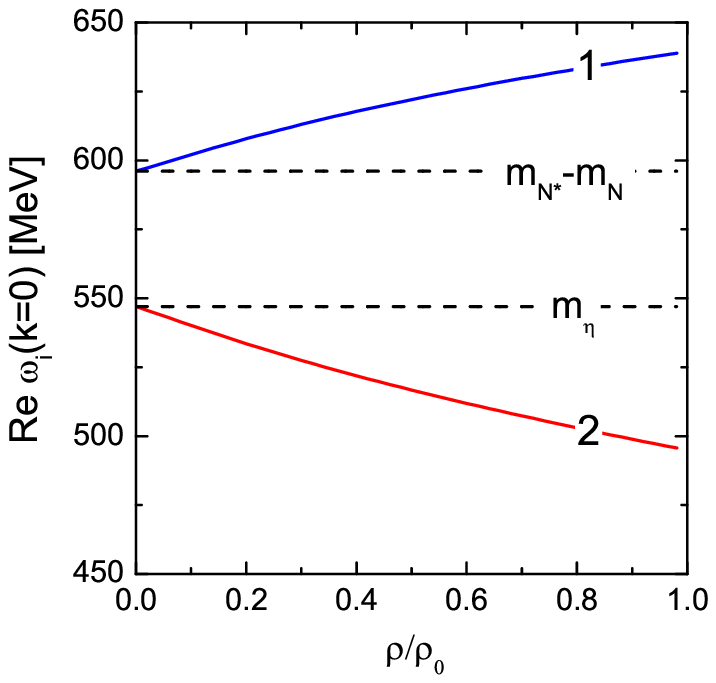}}
\parbox{4.7cm}{\includegraphics[width=4.7cm]{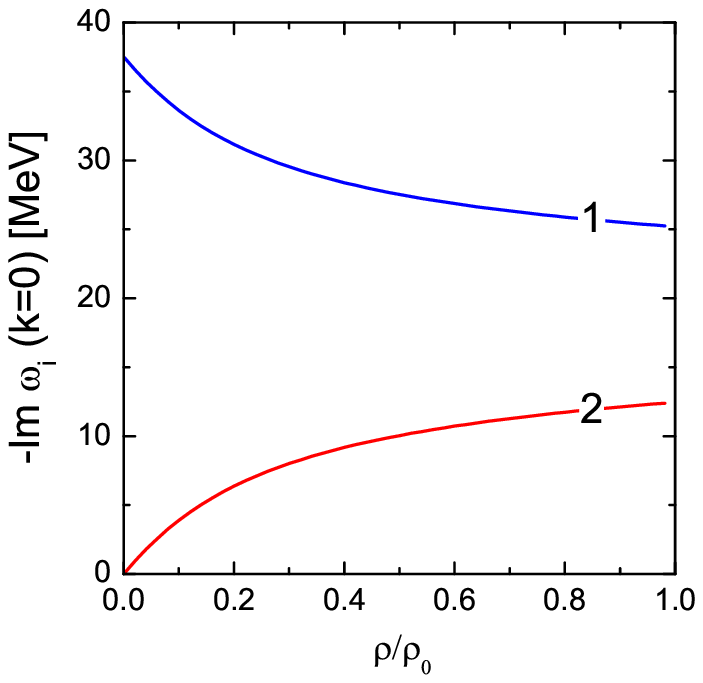}}
\parbox{4.7cm}{\includegraphics[width=4.7cm]{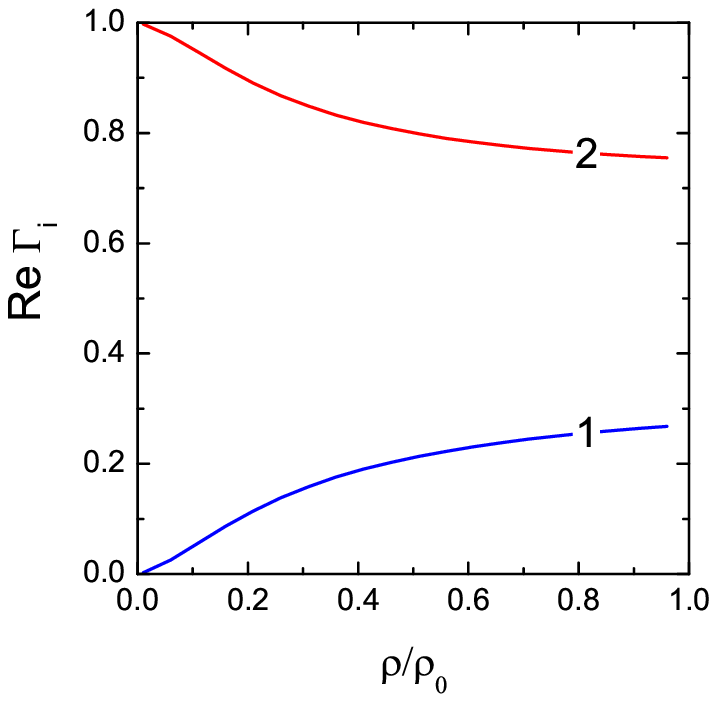}}
}
\caption{\label{fig:solk0C0} The same as in Fig.~\ref{fig:solk0} but calculated with $C=0$.}
\end{figure}

The residua of the poles of the propagator $G_{\eta}(\omega,k)$ at
$\omega = \omega_{i}$ quantify the occupation of each
branch by the eta-meson quantum numbers:
\be
\Gamma_{i}(k;\rho)=2\,m_\eta\, Z_i(k;\rho)\,,\quad
Z_i(k;\rho)= \left. \left(2\om-\frac{\prt \Pi_{\eta}(\omega,k;\rho)}{\prt \om}\right)^{-1}\right |_{\om=\om_i(k)}
\,.\label{occupf}
\ee
Note, it is important to include the imaginary part $\gamma_i$ by
evaluating $Z$-factors. Then they have the following property
$\Im Z_1(k)+\Im Z_2(k)\simeq 0$ and $\Gamma_{1}(k;\rho)+\Gamma_{2}(k;\rho)=1$\,.
The dependence of the occupation factors on the nucleon density
is shown on the right panel in  Fig.~\ref{fig:solk0}.
We see that at $\rho<0.4\, \rho_0$ the eta quantum numbers
populate dominantly branch~2 and for $\rho > 0.4\, \rho_0$ branch~1.

In order to illustrate the effects of the chiral symmetry
restoration we depict in Fig.~\ref{fig:solk0C0} the same eta spectrum parameters as in
Fig.~\ref{fig:solk0} but calculated for $C=0$.
The mass gap of branch 1, $\om_1(k=0)$, increases now with the
increasing density and that of branch 2 decreases half as strong
as for $C=0.2$. The occupation factors show that branch 2 carries
now the dominant part of the eta mesons.

\begin{figure}
\bc
\includegraphics[width=12cm]{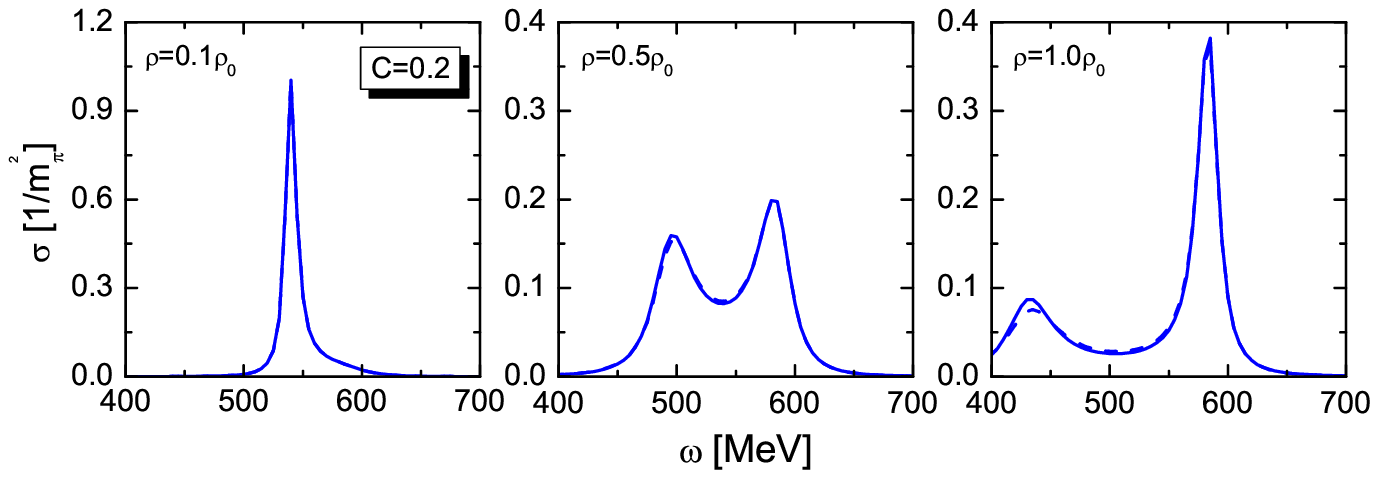}\\
\includegraphics[width=12cm]{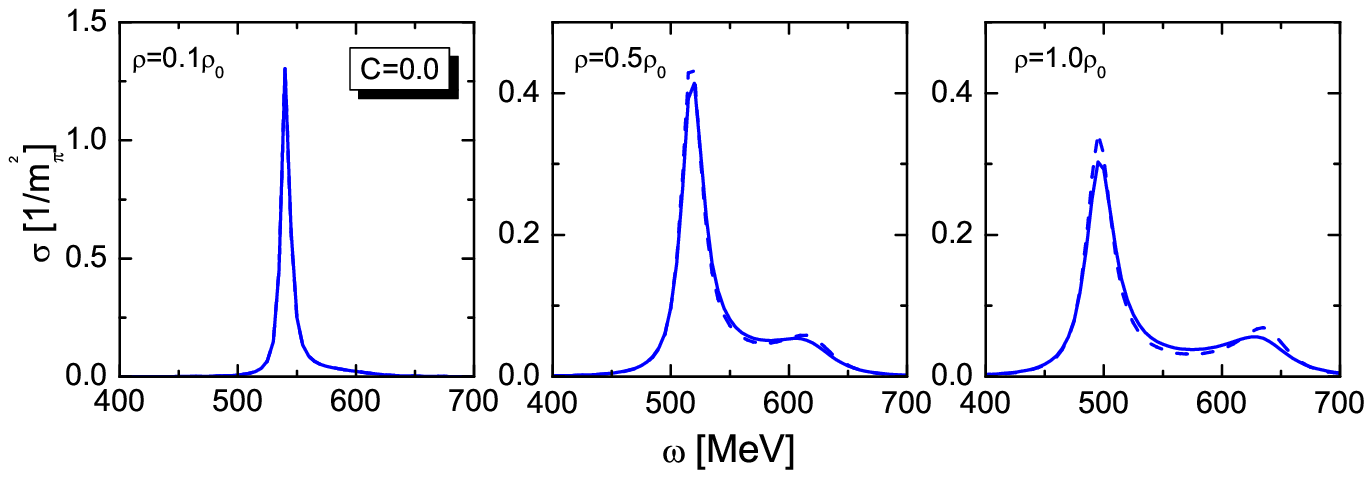}
\ec
\caption{The $\eta$ meson spectral density (\ref{specdens})
in nuclear matter at densities $0.1\rho_0$, $0.5\rho_0$ and $1\rho_0$
for the vanishing momentum.
Dashed lines are calculated with the constant $N^*$ width
$\Gamma_{N^*}=75$~MeV. The solid lines show the results with the
energy and density dependent width shown in
Fig.~\ref{fig:nswidth}.
The upper row is calculated with $C=0.2$ and the lower one
with $C=0$\,.}
\label{fig:spd}
\end{figure}

The spectral function is defined by the imaginary part of
the full propagator (\ref{propeta}):
\begin{equation}
\sigma(\omega, k;\rho) = -\frac{1}{\pi} \Im\, G_{\eta}(\omega,k;\rho)\,.
\label{specdens}
\end{equation}
In Fig.~\ref{fig:spd}
the spectral density
is shown for several values of the nucleon density at the vanishing momentum.
At lower density ($\rho = 0.1 \rho_{0}$) there is one peak, which
corresponds to the eta mesonic mode. At higher densities a two-peak
structure emerges. The second peak comes from the $N^{*}$-$h$ mode which
couples to the eta mode in nuclear medium. The coupling strength
increases as the density increases.
The dashed  curves are calculated with the constant $N^*$ width
$\Gamma_{N^*}=75$~MeV. The solid curves are calculated with the full
energy and density dependent $N^*$ width depicted in
Fig.~\ref{fig:nswidth}.
We note that dashed and solid curves coincide almost perfectly.
For $C=0$ the low-energy peak of the spectral density
is stronger pronounced than for $C=0.2$ though it is located at
somewhat higher energies. It is interesting to see how these
differences in the spectral function reveal themselves in
the finite nuclear systems and in eta production reactions.

\section{Eta meson modes in finite nuclei}

\subsection{Density-averaged spectral function}
In eta production reactions on finite nuclei the spectral function
(\ref{specdens}) is probed at different densities which produce
superposed contributions to the observed final signal. Therefore,
in order to get a feeling on the response of a nucleus to an
external source with the eta-meson quantum numbers its instructive
to study the spectral function averaged over the density
distribution in a nucleus:
\begin{equation}
\overline{\sigma}(\omega, k) = -\frac{4}{A}\int \rmd r\, r^2\, \rho(r)\Im\,
G_{\eta}(\omega,k;\rho(r))\,.
\label{eq:approRF}
\end{equation}
Here $\rho(r)$ is a spatial distribution of the nucleon density
in a  nucleus normalized to the total number of nucleons $A=4\,\pi\int \rmd r\, r^2\, \rho(r)$\,.
\begin{figure}
\centerline{
\includegraphics[width=10cm]{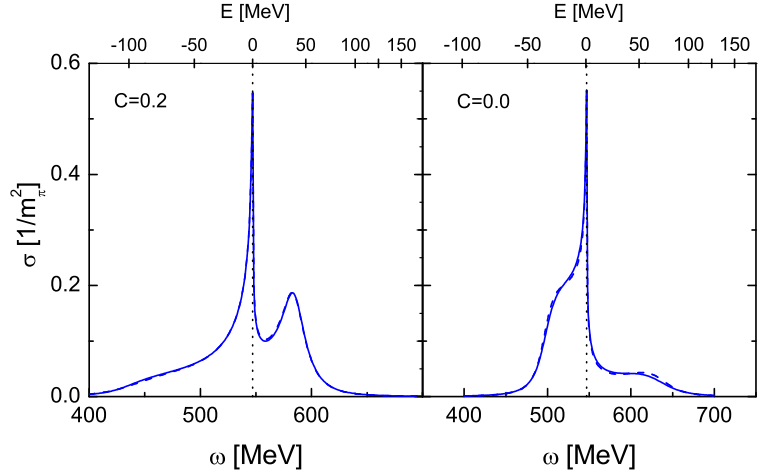}}
\caption{The density-averaged nuclear response function (\ref{eq:approRF}) at zero momentum as a function
of the eta meson energy, $\om$, (lower scale) and the binding energy $E=(\om^2-m_\eta^2)/2\, m_\eta$ (upper scale)
calculated for $C=0.2$ (left panel) and $C=0$ (right panel).
Dashed lines correspond to the constant $N^*$ width and solid
lines to the energy and density dependent width. The vertical
dotted lines mark the bare eta mass value.
}\label{fig:ARF}
\end{figure}
We use the 2-parameter Fermi distribution:
\begin{equation}
\rho(r) =\frac{ \rho_{0} }{1+\exp\left(\frac{r-R}{a}\right)}\,.
\end{equation}
For $^{11}$B, which is the remaining nucleus for eta
production reactions on $^{12}$C, we have
the central density $\rho_{0}= 0.173$ fm$^{-3}$, the
nuclear radius $R= 1.18 A^{1/3}-0.48= 2.14$ fm and the diffuseness
parameter $a=0.5$~fm.

In Fig.~\ref{fig:ARF} we plot the density-averaged nuclear response
functions (\ref{eq:approRF}) with $C=0.0$ and $C=0.2$. Recalling
the in-medium spectral function shown in Fig.~\ref{fig:spd}, we
find the following correspondence between the structures in these
two response functions: the pronounced cusp structure at the eta
production threshold corresponds to lower
densities, when the eta is produced at the surface of the nucleus,
and medium effects are small as seen on the left panel
in Fig.~\ref{fig:spd}. Two bumps --- one in the continuum region
($E>0$) and the other one in the bound regions ($E<0$) --- correspond to the two
modes in the spectrum of the eta meson produced in the nucleus
interior. The difference between the $C=0$ and $C=0.2$ cases
observed in Fig.~\ref{fig:spd} can be clearly seen also in the
density-averaged response function. For $C=0.2$ the bump in
the continuum region is more pronounced and shifts to the higher
energies, whereas the bump in the bound region slides down to
lower energies and becomes smaller. Two modes repel each other
stronger. Therefore, the response strength in the
continuum region larger than that in the bound region can be a signal of
the eta and $N^*-h$ level crossing in the nucleus.

It is worth noting that Eq.~({\ref{eq:approRF}) is not applicable
for calculation of the bound state parameters, since the Green
function calculated in the infinite nuclear matter satisfy the
boundary conditions different from those for the Green function in
a finite nucleus. Therefore, the infinite-matter Green function
does not have correct analytic properties as the actual {\em
in-nucleus} Green function and direct identification of the bumps
in Fig.~\ref{fig:ARF} with the poles of the Green function is not
possible.
We note that identification of the density-averaged spectral
function (\ref{eq:approRF}) with a local approximation of the actual in-nucleus Green
function is not appropriate in regions where the nuclear density
changes suddenly, like on the surface of a nucleus. Therefore, the cusp structures we observe
in Fig.~\ref{fig:ARF} can be only  qualitatively compared to the actual spectra.

\subsection{Bound and resonance states of eta-nucleus system}

In order to understand better the bump structure of the response
function shown in Fig.~\ref{fig:ARF}, we have to study poles of the
eta-meson-nucleus scattering amplitude on the complex energy
plane. The positions of the poles correspond to energies and
widths of bound states of the eta and $N^{*}$-$h$ modes in the
nucleus. In practice, we search for the energy eigenvalues of the Klein-Gordon equation
in the momentum space~\cite{Landau:1982iu} for the eta meson with
the angular momentum $L$ in the nucleus:
\begin{equation}
p^{2} \Psi(p) +
\int_{0}^{\infty} p^{\prime \, 2} dp^{\prime} v_{L}(p,p^{\prime};\omega)
 \Psi(p^{\prime}) = (\om^2-m_\eta^2)\, \Psi(p)\,,
\end{equation}
where $\Psi$ is the wave function of the eta meson,
and $\omega$ is the energy of the eta meson.
The non-relativistic binding energy is defined as
$E=(\om^2-m_\eta^2)/(2\,m_\eta)$.
The potential $v_{L}(p,p^{\prime};\omega)$ in the
momentum space  is given by the Fourier transform of the
energy-dependent self-energy $\Pi_{\eta}(\omega,0;\rho(r))$ evaluated in
Eq.(\ref{polop}) under the local density approximation:
\begin{equation}
v_{L}(p,p^{\prime};\omega) =
\frac{2}{\pi} \int_{0}^{\infty} \rmd r\, r^{2}\, \Pi_{\eta}(\omega,0;\rho(r))
j_{L}(p\,r)\, j_{L}(p^{\prime}\,r)
\end{equation}
with the spherical Bessel function $j_{L}(x)$.
We put $k=0$ since we consider  slow eta excitations created for the
recoilless kinematic conditions.
The boundary conditions can be implemented  more easily by
considering the Green functions in Lippmann-Schwinger equation
rather than the wave function in the Klein-Gordon equation. The
corresponding Lippmann-Schwinger equation written as an
operator equation in the momentum space is
\begin{equation}
\Psi =    G^{(0)}(\om)\, V(\om)\, \Psi\,, \label{eq:LSeq}
\end{equation}
where the Green function $G^{(0)}$ and the potential $V$ are given
in the operator form by $ G^{(0)}(\om)  = (\om^2-m_\eta^2- \hat
p^{2}+i\epsilon)^{-1}$ and $ V(\om) = v_{L}(\hat p, \hat
p^{\prime}; \omega) \hat p^{\prime\, 2}$. The explicit form of
Eq.(\ref{eq:LSeq}) is
\begin{equation}
\Psi(p) - \int_{0}^{\infty} dp^{\prime} \frac{v_{L}(p,p^{\prime}; \omega)\,  p^{\prime\, 2}}{\om^2-m_\eta^2 - p^{\prime\, 2} + i \epsilon} \Psi(p^{\prime}) = 0 \ .
  \label{eq:LSeqIntForm}
\end{equation}
The bound state energies $E$, for which non-trivial solutions
of the Lippmann-Schwinger equation (\ref{eq:LSeq}) exist, follow
from the secular equation
\begin{equation}
   \det \Big( 1- V\big(\sqrt{m_\eta^2+2\,m_\eta\, E}\big)\,
   G^{(0)}\big(\sqrt{m_\eta^2+2\,m_\eta\, E}\big)\Big) = 0 \ .
\end{equation}
Some technical details of the calculations are given in Appendix~\ref{app:A}.

We discuss first the results obtained for $L=0$ and $C=0.2$.
On the first Riemann sheet we find two poles at $E_1=(-88.53-i\,
11.54)$~MeV and $E_2=(-74.13-i\,15.00)$~MeV, which correspond to
the deepest bound $\eta$ state in the $^{11}$B nucleus and its
first radial excitation, respectively. In the response function in
Fig.~\ref{fig:ARF} they manifest themselves as a small bump at
about $-80$~MeV. On the first Riemann sheet there is another pole
at $E=(6.40-i\,0.46)$~MeV. Located above the threshold, this pole
stands for a virtual state, which is not seen in the spectrum as a
resonance bump but causes the cusp at the threshold.
On the second Riemann sheet there is a resonance state at $E=
(61.02-i\,25.82)$~MeV, which shows up as a bump at $\sim 50$~MeV in the
continuum region. Thus we conclude that the structure of the response
function in Fig.~\ref{fig:ARF} (left panel) contains signals of
the two bound, one virtual and one resonance states.
For the $C=0$ case the analytical structure changes. The
attraction weakens and there remains only one bound state at
$(-22.18 -i\,12.88)$~MeV on the first Riemann sheet. The other
former bound state moves up and becomes a resonance at $(3.47 -i\,
54.78)$~MeV on the second Riemann sheet. The other resonance state
on the second Riemann sheet is located now at somewhat smaller energy $(31.91 -i\,54.79)$~MeV.
The virtual state on the first Riemann sheet stays almost put at
$(6.13-i\,0.69)$~MeV.

\section{Formation spectra of $\eta$ mesic nuclei}

In this section we discuss the eta meson spectra in finite nuclei
which can be observed in actual experiments. We consider nucleon
pickup processes of the eta production on a nucleus.
The $(\pi,p)$, (d,$^{3}$He) and $(\gamma,p)$
processes on nuclear targets have been already proposed as formation
reactions of the eta-nucleus systems. The advantage of the nucleon
pickup processes is that it is possible to create the eta
meson in nucleus almost at rest (recoilless production).
The momentum transfer can be reduced to zero by adjusting the incident
particle energy.
For the recoilless production, we have a selection rule for the total
angular momentum of the produced meson and the nucleon-hole state
inside the nucleus. Since at the recoilless kinematics no angular momentum
is brought to the residual nuclear system, the substitutional states are largely enhanced.
As the result  the contributions of the
different angular momenta among the mesonic and nuclear states are
strongly suppressed. For the $^{12}$C target case, the dominant contributions
are provided by the $ (0 s_{1/2})_{p}^{-1} \otimes s_{\eta}$ and $ (0 p_{3/2})_{p}^{-1}
\otimes p_{\eta}$ configurations, in which  $ (n L_{S})_{p}^{-1}
\otimes \ell_{\eta}$ denotes $\eta$-hole configuration with the
$\eta$ meson in the orbit $\ell$ and the proton hole in the
orbit $L$ with the spin $S$ and the principal quantum number $n$,
although the ground state configuration is  $ (0 p_{3/2})_{p}^{-1} \otimes s_{\eta}$,
which has different angular momenta for the $\eta$ and proton-hole states \cite{Hayano:1998sy}.

In the experiments one observes the ejected nucleon in the final states inclusively and obtains the
double differential cross sections $\rmd^2\sigma/(\rmd\Omega \rmd E_N)$ as a functions of the
energy, $E_N$, and the emission angle of the outgoing nucleon. Investigating the shape of the
missing mass spectra, we may hope to extract the in-medium properties of the eta and $N^{*}$-$h$
modes. Observing additional particles in the final state would help to confirm the creation of the
eta and $N^{*}$-$h$ states in the nucleus (see, e.g., proposal~\cite{Sokol91}).

Let us consider the $^{12}{\rm C}(\gamma,p)^{11}{\rm B}$ reaction
as an example. The (d,$^{3}$He) reactions were discussed in
Refs.~\cite{Hayano:1998sy,Jido:2002yb,Nagahiro:2003iv} and the
$(\pi,p)$ will be investigated in Ref.~\cite{Nagahiro}. In the
distorted-wave impulse approximation (DWIA) for the elementary process $\gamma p \rightarrow \eta
p$, the differential cross section is given by the product of the elementary cross section
and the nuclear response function $S(E)$:
\begin{eqnarray}
  \left( \frac{\rmd^{2}\sigma}{\rmd\Omega dE_N}\right)_{A(\gamma,p)(A-1)\otimes \eta}
  =\left(\frac{\rmd\sigma}{\rmd\Omega}\right)_{p(\gamma,p)\eta}^{\rm lab}  \times
S(E_\gamma-E_N) \ . \label{eq:Green}
\end{eqnarray}
where $E_\gamma$ is the incident-photon energy.
In (\ref{eq:Green}) we assume that the two-body process $\gamma p \rightarrow \eta p$
dominates the eta production and the direct production of the
$N^{*}$-$h$ mode by the incident photon via the reaction
$\gamma NN \rightarrow NN^{*}$ is subdominant.

We calculate the nuclear response function $S(E)$ following the
Green function method for the hadrons in nuclei formulated in
Ref.~\cite{GreenFuncMethod}. There the response function is given
by the transition amplitudes $T_{f}({\bf r})$ of $\gamma
\rightarrow p^{-1} p \eta$ at the position $\bf r$ and the Green
function $G(E;{\bf r}, {\bf r}^{\prime})$ of the eta meson in the
residual nucleus:
\begin{eqnarray}
  S(E) = -\frac{1}{\pi} {\rm Im} \int \rmd{\bf r}\, \rmd{\bf r}^{\prime}
  \sum_{f} { T}_f^{\dagger}({\bf r}) G(E-\Delta E_{j_p}; {\bf r}, {\bf r}^{\prime}) { T}_f({\bf r}^{\prime}) \ , \label{eq:resfunc}
\end{eqnarray}
where the summation is taken over the all possible finial states and
$\Delta E_{j_p}$ is the hole separation energy measured from the ground state of the daughter nucleus.
After the summation of the nucleon spins the transition amplitude
$T_{f}$ is given by
\begin{equation}
  {T}_f({\bf r}) = \chi^{*}_{f}({\bf r})\, \xi^{*}_{1/2,m_{s}}
  \left[Y_{l_{\eta}}^{*}(\hat r)\otimes \psi_{j_{p}}({\bf r})\right]_{JM} \chi_{i}({\bf r})
  \label{eq:tranamp}
\end{equation}
with the proton--hole wave function $\psi_{j_p}$, the distorted
waves $\chi_{i}$ and $\chi_{j}$ of the incident photon and the
ejected proton, the eta meson angular wave function
$Y_{l_{\eta}}(\hat r)$ and the spin wave function $\xi_{1/2,m_s}$
of the ejected proton. $JM$ denotes the total angular momentum of
the $\eta$ and proton-hole system.
The Green function $G(E; {\bf r}, {\bf r^{\prime}})$ is written
formally as
\begin{equation}
   G(E; {\bf r}, {\bf r^{\prime}}) = \langle p^{-1} | \phi_{\eta}({\bf r})
\frac{1}{E-H_{\eta}+i\epsilon} \phi_{\eta}^{\dagger}({\bf r^{\prime}}) |
p^{-1} \rangle
\label{eq:Greenfunction}
\end{equation}
with the Hamiltonian containing the eta meson optical potential
\begin{equation}
   U_{\eta}(\omega,r) = \frac{1}{2\,m_\eta}
   \Pi_{\eta}(\omega;\rho(r))\,.
   \label{eq:optpot}
\end{equation}
In (\ref{eq:Greenfunction}) $\phi^{\dagger}_{\eta}$ is the meson creation operator and
$|p^{-1}\rangle$ is the residual nucleus state with a proton hole.
The Green function $G(E; {\bf r}, {\bf r^{\prime}})$ given in
Eq.(\ref{eq:Greenfunction}) is calculated by solving the
Klein-Gordon (or Schr\"odinger) equation with appropriate boundary
conditions. In general, the Green function can depend on the
nuclear states and give transitions between different states. Here
we have assumed that these state dependence and transitions are
negligibly small.

The density averaged spectral function (\ref{eq:approRF}) is related to the actual response
function (\ref{eq:resfunc}) in the following way: since
at the recoilless production the eta meson created in the nucleus moves slowly,
it cannot propagate a long distance being absorbed rather soon.
In this case the in-nucleus Green function $G(E; {\bf r}, {\bf
r^{\prime}})$ given in Eq.(\ref{eq:Greenfunction}) can be approximated as
\begin{equation}
   G(E; {\bf r}, {\bf r^{\prime}}) \simeq \delta({\bf r}-{\bf r^{\prime}}) G_{\eta}(E;\rho(r)),
\end{equation}
where $G_{\eta}(E;\rho(r))$ is the Green function for infinite nuclear matter obtained in the
local density approximation. Neglecting the distortion effects on the incident photon and the
ejected proton in the transition amplitude $T_{f}({\bf r})$ in Eq.(\ref{eq:tranamp}) and replacing
the product of the hole wave-functions, $\psi_{j_{p}}^{\dagger}({\bf r})\psi_{j_{p}}({\bf r})$ by
the density distribution $\rho(r)$ after the summation of the final states, we obtain an
approximate response function
\begin{equation}
   S(E) \propto \int dr  r^{2}  \rho(r)\,  {\Im}\, G_{\eta}(E;\rho(r))\,.
\end{equation}
This is the density weighted in-medium Green function introduced in (\ref{eq:approRF}) except for
the normalization. According to the above argument, the approximate spectral function is close to
the in-nucleus spectral function obtained by solving the Klein-Gordon equation, if the local
density approximation is good enough. In the proceeding calculations of the formation spectra of  eta
mesic nuclei we use the complete in-nucleus Green function (\ref{eq:Greenfunction}).

\begin{figure}
\centerline{\includegraphics[width=12cm]{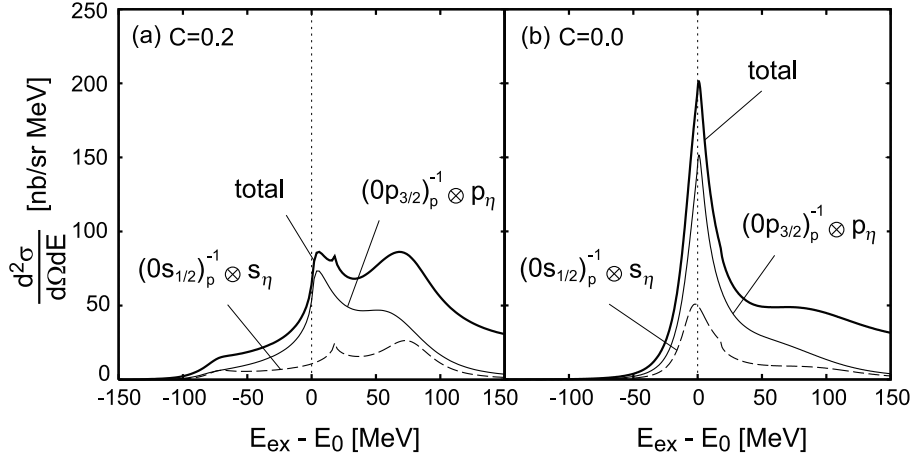}} \caption{
Formation spectra of the eta meson in the $(\gamma,p)$ reaction
on the $^{12}$C target at $E_{\gamma}=950$ MeV as a function of the energy of the proton, $E_{\rm ex}$,
emitted at zero angle. The proton energy is
counted from the eta production threshold $E_0$.
The left and right panels are the formation spectra calculated in the chiral
doublet model (the naive model) with $C=0.2$ and $C=0.0$,
respectively. \label{fig:spec}}
\end{figure}

The calculated formation spectra of the eta meson in the $(\gamma,p)$ reaction with the $^{12}$C
target are shown in Fig.~\ref{fig:spec}. The incident photon energy $E_\gamma=950$~MeV is chosen so
that the recoilless condition is satisfied for the bare eta mass. We plot the missing mass spectra
in the energy range wider than those shown in Refs.~\cite{Jido:2002yb,Nagahiro:2003iv,Nagahiro:2005gf}
to see the global structure of the spectra and to be able to identify the bound and resonance
states discussed in the previous section.
The energy of the emitted  proton is counted from the creation threshold of the vacuum eta mass.
The left (right) side of the threshold corresponds to the bound (continuum) energy region.
We have calculated the sum of all contributions from the possible configurations of the
proton-hole states, $s_{1/2}$ and $p_{3/2}$, and the eta states with $l=0$ to $6$.
The total spectra are dominated by two contributions
from the $ (0 s_{1/2})_{p}^{-1} \otimes s_{\eta}$ and $ (0 p_{3/2})_{p}^{-1}
\otimes p_{\eta}$ configurations due to the selection rule at the recoilless
kinematics.
The separation energies of the $s$ state protons are taken into account in the spectra. The
formation spectra are discussed in more details in
Refs.~\cite{Jido:2002yb,Nagahiro:2003iv,Nagahiro:2005gf}.

The missing mass spectrum calculated in the chiral doublet model with $C=0.2$ and shown on the
left panel in Fig.\ref{fig:spec} has the richer structure than the spectrum obtained with $C=0.0$,
when no level crossing takes place. We compare the sub-component $ (0 s_{1/2})^{-1}_{p} \otimes
s_{\eta}$ in Fig.\ref{fig:spec} with the response function plotted in Fig.~\ref{fig:ARF}. The
centrifugal potentials are not taken into account in the latter function. We find that still the
approximated response function reproduces well the bump structure of the spectrum obtained in the
full calculation with the actual Green function. 
(Exception is the eta threshold energies, when the
contributions from the eta meson on the surface of the residual nucleus are dominated.) This is
because the distortion factors have smooth energy dependence and the initial photon is less
distorted. Thus, the conclusions drawn in the previous section for the approximated spectra hold
also for the actual formation spectra. Namely, the bump structures correspond to the bound and
resonance states. For the $C=0.2$ case, we have found the two bound states and one resonance state.
They correspond to the small bump around $E=-80$ MeV and the pronounced peak in $E=80$ MeV,
respectively.

We conclude that the more pronounced bump structure in the continuum energy region can be a signal
of the level crossing.
Note that the recoilless kinematics largely suppresses
the second bound state (the radial excited state) in the spectrum, since there
are no proton-hole states in $^{11}B$ to satisfy the selection rule with the second
$s$ state of the eta meson.

\section{Conclusion}

We have investigated the eta-nucleus system, paying attention to
the $N(1535)$--nucleon-hole mode in nuclear medium. We have found
that the $N^{*}$--$h$ contribution is important for the eta meson
in nuclear matter and induces strong energy dependence of the eta
self-energy. We have also shown that reduction of the difference
between the masses of $N^{*}$ and $N$ as a result of the medium
effects leads to level crossing of the eta and $N^{*}$--$h$ modes,
and modification of the eta spectral function. The change of the
spectral function is clearly seen in formation spectra of eta
nuclei.

\section*{Acknowledgments}
E.E.K. thanks the Yukawa Institute for hospitality.
This work is supported by the Grant-in-Aid for the 21st Century
COE "Center for Diversity and Universality in Physics" from the
Ministry of Education, Culture, Sports, Science and Technology
(MEXT) of Japan. 
The work of D.J. was also partially supported by the
Grant for Scientific Research (No.\ 18042001) from MEXT.
The work of E.E.K. was supported in
part by the US Department of Energy under contract No. DE-FG02-87ER40328.
The work of H.N. was supported by Japan Society for the
Promotion of Science (No.~18-8661).
This work is part of Yukawa International Program for Quark-Hadron Sciences.


\appendix
\section{Appendix}\label{app:A}

The numerical solution of the equation (\ref{eq:LSeqIntForm}) is performed by the method suggested in
Ref.~\cite{Landau:1982iu}. The integral equation is written as a matrix equation
in momentum space by replacing momentum integral with a sum over
$N$ grid points:
\begin{equation}
   \Psi_{m} = \sum_{n=1}^{N} M_{mn} \, \Psi_{n}
\end{equation}
with
\begin{equation}
   M_{mn} = V_{mn}(\om)\, G_{n}^{(0)}(\om)\, W_{n}\,,
    \label{eq:LSmat}
\end{equation}
where $W_{n}$ is a weight function for the discretization of the integral and
\begin{eqnarray}
 V_{mn}(\om) &=&  v_{L}(p_{m}, p_{n}; \omega)\, p^{2}_{n}\,, \\
 G^{(0)}_{n}(\om) &=&  \frac{1}{ \om^2-m_\eta^2- p^{2}_{n}+i\epsilon}\,. \label{eq:green_n}
\end{eqnarray}
At the end we search for zeros of the determinant of the $N \times N$ matrix:
\begin{equation}
   \det(1- M) = 0 \label{eq:detzero}
\end{equation}
for the complex binding energy $E=(\om^2-m_\eta^2)/(2\,
m_\eta)$\,.

In this method, one can choose properly the boundary conditions of
the scattering equation. For the first Riemann sheet
solutions, we just solve Eq.(\ref{eq:detzero}) for a complex $E$.
For the second Riemann sheet, where we can obtain
resonance solution, one should treat $i\epsilon$ in the Green
function $G^{(0)}$ given in Eq.(\ref{eq:green_n}) before going to
the complex plane. This can be done as follows. The integral in
the left hand side of Eq.(\ref{eq:LSeqIntForm}) can be written in
terms of principle value integral as
\begin{eqnarray}
\Psi(p)=
P \int_{0}^{\infty} dp^{\prime} \frac{v_{L}(p,p^{\prime}; \omega)\,  p^{\prime\, 2}}
{\om^2-m_\eta^2 - p^{\prime\, 2}} \Psi(p^{\prime})
- i\frac12\,\pi k v_{L}(p,k;\omega) \Psi(k)\,,
\end{eqnarray}
where $k$ is a complex momentum   $k^{2}=\om^2-m_\eta^2 =2\,
m_\eta\, E$ on the second Riemann sheet. As discussed in
Ref.~\cite{Haftel:1970fk}, these terms are implemented to the
matrix formulation by extending the matrix $M$ in (\ref{eq:detzero}) and defining an extra
grid point $p_{N+1}=k$ as
\begin{eqnarray}
   M_{N+1,n} &=&  v_{L}(k,p_{n};\omega) \frac{W_{n} p_{n}^{2}}{k^{2}-p_{n}^{2}}\,, \\
   M_{n,N+1} &=&  v_{L}(p_{n}, k; \omega) \left( \sum_{m=1}^{N} \frac{-W_{m}k^{2}}{k^{2}-p_{m}^{2}}   - i \frac12\,\pi k \right)\,,  \\
   M_{N+1,N+1} &=& v_{L}(k,k;\omega) \left( \sum_{m=1}^{N} \frac{-W_{m}k^{2}}{k^{2}-p_{m}^{2}}   - i \frac12\pi k
   \right)\,.
\end{eqnarray}

\end{document}